\documentclass{iau} 
\usepackage{graphicx}

\title[Comparison of Solar White-light Flares and Superflares] 
{Statistical Study of Solar White-light Flares and Comparison with Superflares on Solar-type Stars}

\author[Namekata et al.]   
{Kosuke Namekata$^{1, *}$,
Takahito Sakaue$^{1}$,
Kyoko Watanabe$^{2}$,
Ayumi Asai$^{3,4}$,
Hiroyuki Maehara$^{1}$,
Yuta Notsu$^{1}$,
Shota Notsu$^{1}$,
Satoshi Honda$^{5}$,
Takako T. Ishii$^{3}$,
Kai Ikuta$^{1}$,
Daisaku Nogami$^{1}$,
\and
Kazunari Shibata$^{3,4}$}

\affiliation{
$^*$ email: {\tt namekata@kusastro.kyoto-u.ac.jp}  \\

$^1$Department of Astronomy, Kyoto University, Kitashirakawa-Oiwake-cho, Sakyo-ku, Kyoto 606-8502, Japan\\
$^2$National Defense Academy of Japan, 1-10-20 Hashirimizu, Yokosuka, 239-8686, Japan\\
$^3$Kwasan and Hida Observatories, Kyoto University, Yamashina, Kyoto 607-8471, Japan\\
$^4$Unit of Synergetic Studies for Space, Kyoto University, Yamashina, Kyoto 607-8471, Japan\\
$^5$Nishi-Harima Astronomical Observatory, Center for Astronomy, University of Hyogo, Sayo-cho, Sayo-gun, Hyogo, 679-5313, Japan}

\pubyear{2018}
\volume{340}  
\jname{Statistical Study of Solar White-light Flares and Comparison with Superflares on Solar-type Stars}
\editors{}
\begin{document}

\maketitle

\begin{abstract}

Recently, many superflares on solar-type stars were discovered as white-light flares (WLFs). 
A correlation between the energies (E) and durations (t) of superflares is derived as $t\propto E^{0.39}$, and this can be theoretically explained by magnetic reconnection ($t\propto E^{1/3}$).
In this study, we carried out a statistical research on 50 solar WLFs with \textit{SDO}/HMI to examine the t-E relation. 
As a result, the t-E relation on solar WLFs ($t\propto E^{0.38}$) is quite similar stellar superflares,
but the durations of stellar superflares are much shorter than those extrapolated from solar WLFs. 
We present the following two interpretations; (1) in solar flares, the cooling timescale of WL emission may be longer than the reconnection one, and the decay time can be determined by the cooling timescale; (2) the distribution can be understood by applying a scaling law $t\propto E^{1/3}B^{-5/3}$ derived from the magnetic reconnection theory. 

\keywords{solar flares, stellar flares, magnetic reconnection}
\end{abstract}

\firstsection 
\section{Introduction}
Solar flares are abrupt brightenings on the solar surface. 
During flares, magnetic energy is believed to be converted to kinetic and thermal energies through the magnetic reconnection in the corona (e.g., \cite[Shibata \& Magara 2011]{2011LRSP....8....6S}).
In the standard scenario, the released energies are transported to lower atmosphere to emit electromagnetic waves across wide wavelength ranges.
Flares in the visible continuum are particularly called white-light flares (WLFs).
Although it has passed about 150 years since the first solar WLF was observed, the emission mechanism is not well understood (\cite[Heinzel et al. 2017]{2017ApJ...847...48H}).

Recently, many stellar WLFs were found by space-based optical telescopes.
Interestingly, they also discovered large WLFs, called ``superflares'', on solar-type stars (G-type main sequence stars) whose energies ($10^{33-36}$ erg) are 10--10,000 times larger than those of the maximum solar flares ($\sim10^{32}$ erg;  \cite[Maehara et al. 2012, Shibayama et al. 2013]{2012Natur.485..478M,2013ApJS..209....5S}).
The occurrence frequency of flares and spots are universally expressed with the same power-law relation among the Sun and superflare stars (\cite[Shibayama et al. 2013, Maehara et al. 2017]{2013ApJS..209....5S,2017PASJ...69...41M}).
The released energies through solar flares and superflares are found to be comparable with the magnetic energies stored around the spots (\cite[Notsu et al. 2013]{2013ApJ...771..127N}).
This indicates that superflares are also phenomena where magnetic energies are released.

\cite{2015EP&S...67...59M} reported that there is a correlation between the energies radiated in WL ($E$) and durations ($\tau$) of superflares: $\tau \propto E^{0.39}$.
It is consistent with those of solar flares observed with hard/soft X-rays: $\tau \propto E^{0.2-0.33}$ (\cite[Veronig et al. 2002, Christe et al. 2008]{2002A&A...382.1070V,2008ApJ...677.1385C}).
Moreover, the observed power law relations can be explained by the magnetically driven energy release (magnetic reconnection).
In detail, flare energy is expressed as a function of magnetic field strength ($B$) and length scale ($L$) of flares: $E \sim fE_{\rm mag}\sim fB^2L^3$ ($f$ is a filling factor).
Duration is thought to be comparable to the reconnection time scale ($\tau_{\rm rec}$): $\tau \sim \tau_{\rm rec} \sim L/v_{A}/M_A$ ($v_A$ is the \textit{Alfv\'en} velocity, $M_A$ is the dimensionless reconnection rate $\sim$ 0.1-0.01).
Assuming that stellar properties ($B$ and $v_A$) are not so different among the same spectral-type stars (solar-type stars), the values of both $E$ and $\tau$ are determined by the length scale ($L$).
On the basis of this assumption, the relation between $E$ and $\tau$ can be derived by deleting $L$ from the above equations:
\begin{eqnarray}\label{eq:3}
\tau \propto E^{1/3}.
\end{eqnarray}
This similarity between the theory and observations indirectly indicates that solar and stellar flares can be explained by the magnetic reconnection.  
In this study, we carried out a statistical research on 50 solar WLFs and compared the $E$--$\tau$ relations in the WL wavelength range, aiming to confirm the above expectation that solar and stellar flares can be universally explained by the same theoretical relation (Equation \ref{eq:3}).

\section{Analysis}
We carried out statistical analyses of temporal variations of 50 solar WLFs observed by \textit{Solar Dynamics Observatory} (\textit{SDO})/Helioseismic and Magnetic Imager (HMI) in the continuum channel with 45 sec cadence.
The catalog contains M to X class flares on 2011--2015, which is listed in \cite{2017ApJ...851...91N}.
It is necessary to carefully subtract the background trend since the emission of solar WLFs is difficult to detect due to much lower contrast to the photosphere. 
We identified the WL emissions inside the region with strong HXR emissions observed by \textit{RHESSI} (\textit{Reuven Ramaty High Energy Solar Spectroscopic Imager;}).
The radiated energies and decay times were calculated from the extracted light curves of WLFs. 
We measured the duration as e-folding decay time, and calculated the WL energy by assuming that it is radiated by $T_{\rm flare}=$10,000 K blackbody.

\section{Result \& Discussion}

\begin{figure}[htbp]
\begin{center}
 \includegraphics[width=3.4in]{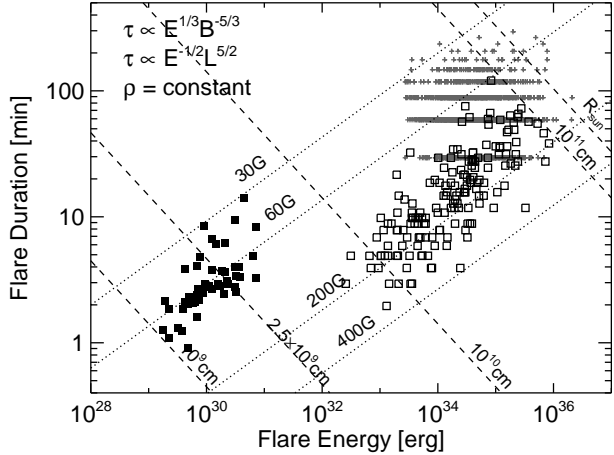} 
 \caption{Comparison between flare radiated energies and durations taken from \cite[Namekata et al. (2017)]{2017ApJ...851...91N}. 
 Filled squares are solar white-light flares. 
 Open squares and crosses are stellar superflares detected by Kepler 1 min and 30 min cadence data.
 Dotted and dashed lines are theoretical scaling law derived on the basis of magnetic reconnection theory (Equation \ref{eq:nam}).}
   \label{fig1}
\end{center}
\end{figure}

The relations between flare energies and durations have been found to be universal among solar X-ray flares ($\tau \propto E^{0.2-0.33}$) and stellar WLFs ($\tau \propto E^{0.39}$), and the relation well matches the theoretical relation consistent with magnetic reconnection ($\tau \propto E^{1/3}$).
Our result also showed that the relation of solar WLFs ($\tau \propto E^{0.38\pm 0.06}$) well matches these previous studies. 
This consistency supports the suggestion that both solar and stellar flares are caused by the magnetic reconnection. 
However, we also found that solar flares and stellar superflares were not on a same line though the power-law indexes are the same ($\tau \propto E^{1/3}$), and the durations of superflares are one order of magnitude shorter than extrapolated from solar flares (Figure \ref{fig1}).
This discrepancy indicates that solar and stellar WLFs cannot be simply explained by the relation derived by \cite{2015EP&S...67...59M}.
We propose two interpretations to explain such a discrepancy.

\subsection{Cooling Effect}
The first interpretation of the $E$-$\tau$ diagram is that the difference between solar and stellar flares is related to the properties unique to WLFs, especially cooling effect.
Solar observations show that solar WLFs have long ($\sim$500 sec) decay components and the timescales correspond to coronal cooling timescales (e.g., \cite[Kawate et al. 2016]{2016ApJ...833...50K}).
We should note that it is not confirmed whether such decay time originates in the cooling timescale or the long lasting reconnection.
However, if it does correspond to the cooling timescale, the 500 sec cooling time ($t_{\rm cool}$) is not negligible compared to the reconnection timescale ($t_{\rm rec}$) and could elongate the decay time of WLFs.
Therefore, there is a possibility that the decay time of solar WLFs (below $\sim10^{32}$ erg) are elongated by cooling effect because $t_{\rm cool} >> \it t_{\rm rec}$, but those of superflares (above $\sim10^{33}$ erg) are not elongated because $t_{\rm cool} << \it t_{\rm rec}$, which can result in the observed $E$--$\tau$ discrepancy.

\subsection{Magnetic Field Strength}
We present here the second interpretation of the observed $E$--$\tau$ diagram.
\cite{2015EP&S...67...59M} derived the theoretical scaling law ($\tau \propto E^{1/3}$) assuming that the Alfv\'{e}n velocity ($v_A=B/\sqrt{4\pi\rho}$) is constant for each flare on solar-type stars.
When considering the dependence on Alfv\'{e}n velocity, the scaling law can be expressed as follows:
\begin{eqnarray}\label{eq:mae}
\tau \propto E^{1/3}B^{-5/3}\rho^{1/2}.
\end{eqnarray} 
On the basis of this scaling law, the one order of magnitude shorter durations of superflares can be understood by (1) two orders of magnitude lower coronal density of superflares, or (2) about a factor of 2--4 stronger coronal magnetic field strength of superflares than that of solar flares. 
The former possibility is less likely because superflare stars are rapidly rotating ones which are expected to have higher coronal densities based on the large emission measures of the X-ray intensity (e.g., \cite[Wright et al. 2011]{2011ApJ...743...48W}).
On the other hand, the latter well accounts for the $E$--$\tau$ distributions without any contradiction with observations that superflare stars show high magnetic activities (e.g., \cite[Notsu et al. 2015]{2015PASJ...67...33N}).
On the basis of such a scaling relation, we proposed that the discrepancy can be caused by the strong coronal magnetic field strength of superflares.
By assuming pre-flare coronal density is a constant value, the following scaling laws can be derived:
\begin{eqnarray}\label{eq:nam}
\tau \propto E^{1/3}B^{-5/3}, \tau \propto E^{-1/2}L^{5/2}.\label{eq:nam2}
\end{eqnarray}
To simply determine the coefficients, we observationally measured the average values of $B$ and $L$ on the basis of the method introduced by \cite{2017PASJ...69....7N}, and the coefficients can be obtained as $B_0=57$ G, $L_0=2.4\times 10^9$ cm, $\tau_0=3.5$ min and $E_0=1.5\times10^{30}$ erg from the average among the solar flares in our catalogue.
On the basis of such values, we applied the scaling laws to the observed $E$--$\tau$ diagram as in Figure \ref{fig1}, and found that solar flares and stellar flares have coronal magnetic field strength of 30--400 G.
This range is roughly comparable to the observed values of solar and stellar flares (e.g., \cite[Shibata \& Yokoyama 2002]{2002ApJ...577..422S}).
It is also reasonable that the estimated loop length of superflares ($10^{10-11}$ cm) is less than the solar diameter ($1.4\times10^{11}$ cm).
According to the scaling law, superflares observed with Kepler data have 2--4 times stronger coronal magnetic field strength than solar flares.

\section{Summary}

We conducted a statistical research on solar WLFs, and compared the relation of flare energy ($E$) and duration ($\tau$) with those of superflares on solar-type stars, aiming to understand the energy release of superflares by the magnetic reconnection theory. 
The results show that superflares have one order of magnitude shorter durations than those extrapolated from the power-law relation of the obtained solar WLFs.
This discrepancy may have a potential to understand the detailed energy release mechanism of superflares as well as properties of the unsolved origin of WL emissions.

We proposed the following two physical interpretations on the $E$--$\tau$ diagram.
(1) In the case of solar flares, the reconnection timescale is shorter than the cooling timescale of white light, and the decay time is determined by the cooling timescale.
(2) The distribution can be understood by a scaling law ($\tau \propto E^{1/3}B^{-5/3}$) obtained from the magnetic reconnection, and the coronal magnetic fields of the observed superflares are 2-4 times stronger than those of solar flares.
The scaling laws can predict the unresolved stellar parameters, the magnetic field strength and loop length.
This would be helpful for investigations on stellar properties in the future photometric observations (e.g., $TESS$).

\end{document}